\documentclass[cameraready]{Interspeech}

\title{What Do Neural Networks Learn for TDOA Estimation?\\A Cross-Architecture Probing Study}

\author{Yaozhong}{Kang}
\author{Jiang}{Wang}
\author{Runwu}{Shi}
\author{Takeshi}{Ashizawa}
\author{Benjamin}{Yen}
\renewcommand{\authorsep}{, and }
\author{Kazuhiro}{Nakadai}
\address{
    Department of Systems and Control Engineering, Institute of Science Tokyo, Japan
}
\email{\{kangyaozhong, wangjiang, shirunwu, ashizawa, benjamin, nakadai\}@ra.sc.eng.isct.ac.jp}

\keywords{TDOA estimation, GCC-PHAT, representation analysis, probing, neural network interpretability}

\usepackage{multirow}
\usepackage{float}

\newcommand{\crossre}{\mathrm{cross_{re}}}

\newcommand{\phatcos}{\mathrm{phat_{cos}}}
\newcommand{\phatsin}{\mathrm{phat_{sin}}}

\newcommand{\blfootnote}[1]{\begingroup\renewcommand{\thefootnote}{}\footnotetext{\raggedright #1}\endgroup}

\begin{document}
 
\maketitle
\blfootnote{Code: \url{https://github.com/york1to/cross-power-is-all-you-need}}

\begin{abstract}
Neural networks outperform classical GCC-PHAT for Time-Difference-of-Arrival (TDOA) estimation in noise and reverberation, yet their internal strategy remains unexplored. To uncover it, we turn GCC-PHAT's mathematical steps into diagnostic targets, probing hidden layers of three architectures (MLP, CNN, Transformer) and complementing with gradient attribution and causal frequency masking. We find that cross-power computation consistently emerges across all architectures and conditions, while PHAT whitening, the defining step of GCC-PHAT, fails to emerge. Instead, networks learn a magnitude-aware frequency weighting that preserves per-frequency reliability information discarded by PHAT. This makes PHAT an information bottleneck: removing it from both classical and neural GCC pipelines improves performance under additive noise. On real-world reverberant data, PHAT remains the best classical weighting, but end-to-end networks achieve lower error by learning data-adaptive weighting.
\end{abstract}
\section{Introduction}
\label{sec:intro}

Time-difference-of-arrival (TDOA) estimation is a core primitive for sound source localization with microphone arrays, with applications from hearing aids to meeting transcription and mobile robotic platforms~\cite{grumiaux2022survey, wang2025observability}. The dominant classical method, generalized cross-correlation with phase transform (GCC-PHAT)~\cite{knapp1976generalized}, estimates delay in three steps: (i)~forming the cross-power spectrum of two microphone signals, (ii)~applying PHAT ``whitening'' to normalize each frequency bin to unit magnitude, and (iii)~taking the inverse Fourier transform and selecting the peak. The whitening step is well-motivated under spatially white noise at high SNR~\cite{zhang2008phat}, yet it degrades under colored noise and behaves unpredictably under reverberation~\cite{ianniello1982time, champagne1996performance}.

\begin{figure}[t]
  \centering
  \includegraphics[width=0.74\columnwidth]{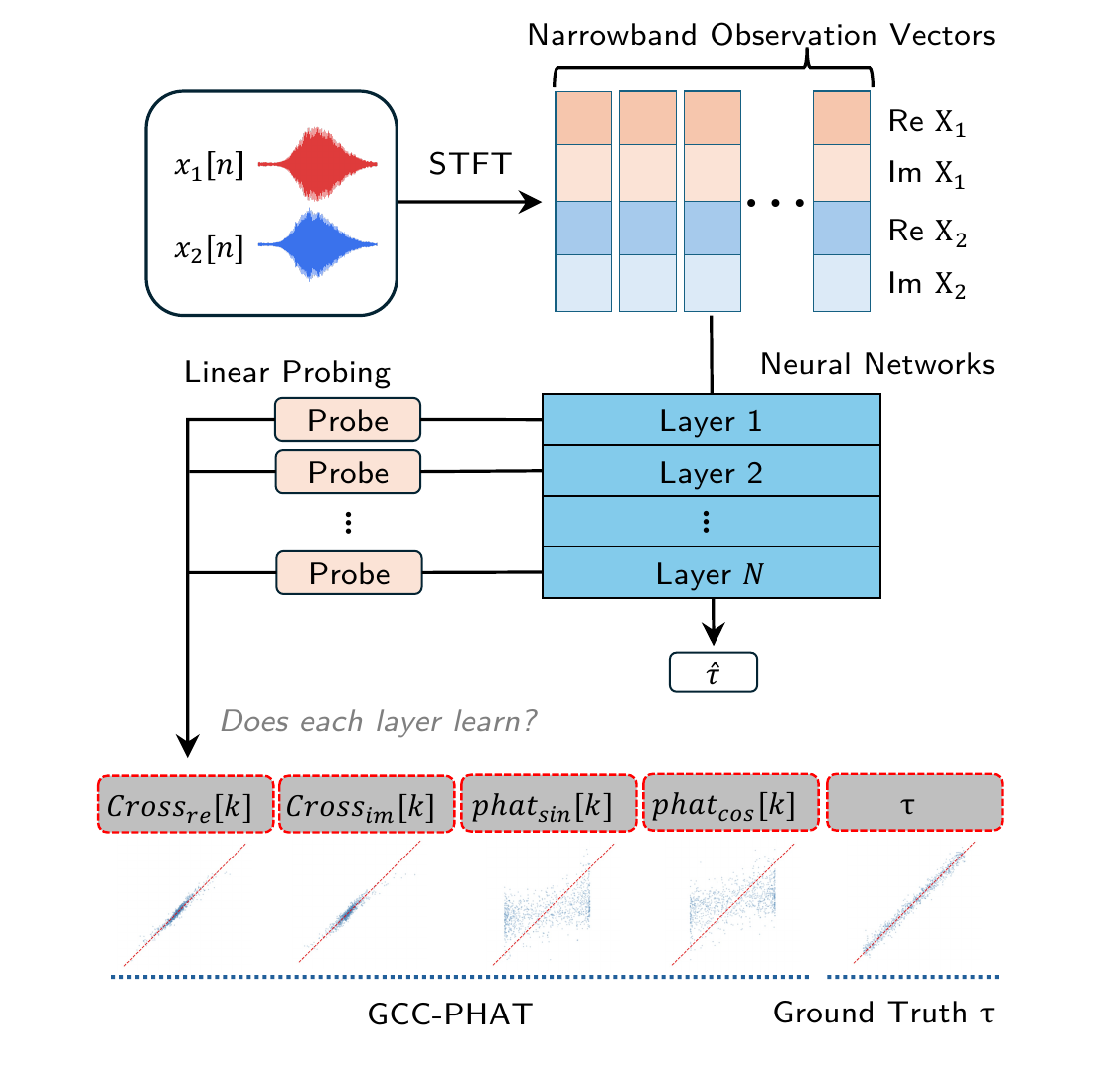}
  \caption{\textbf{Probing framework overview.} Dual-channel signals are transformed via STFT into narrowband observation vectors, processed by one of three network architectures, and probed at each layer for GCC-PHAT intermediate variables: cross-power, PHAT whitening, and the learned magnitude-aware weighting alternative.}
  \label{fig:framework}
\end{figure}

To overcome these limitations, modern sound source localization systems increasingly rely on neural networks, either compact models that learn spectral weighting in the frequency domain~\cite{pertila2019tdoa, salvati21_interspeech, berg22_interspeech}, or large-scale architectures that jointly estimate source location and sound event type~\cite{politis2021overview, hu2025pseldnets}. For determinism and computational efficiency, many practical systems adopt a hybrid design that combines classical signal processing with a neural back-end~\cite{salvati21_interspeech, berg22_interspeech}. However, fixed signal-processing steps may discard information that the neural component could otherwise exploit. Understanding \emph{what} neural networks learn internally is critical for principled design of such hybrid pipelines, yet this question remains largely unexplored.

We address this question by turning GCC-PHAT from a baseline into a diagnostic tool. Specifically, we apply representation probing~\cite{alain2017understanding,belinkov2022probing, conneau2018probing}, which trains a simple linear model on hidden-layer activations to predict a target variable; a high linear decodability indicates the variable is learned in the representation~\cite{voita2020information}. This methodology has revealed interpretable structure in language models~\cite{belinkov2022probing, conneau2018probing}, mathematical reasoning~\cite{nanda2023progress}, and speech representations~\cite{cho2023evidence, choi2024phonetic}. GCC-PHAT is especially suited for probing because each algorithmic step is mathematically specified, yielding exact ground-truth targets. Unlike algorithm unrolling~\cite{monga2021algorithm}, which constrains network architecture to follow a known algorithm, we keep architectures unconstrained and analyze what computations emerge from data-driven learning. To disentangle per-frequency and cross-frequency computation, we study three architectures with systematically different frequency connectivity (per-bin MLP, 1D-CNN, and Transformer), and complement probing with gradient-based attribution~\cite{simonyan2014deep} and causal frequency-masking intervention~\cite{zeiler2014visualizing}.

\textit{First}, we find that cross-power computation consistently emerges across all architectures and conditions, while PHAT whitening fails to emerge in any decodable form. \textit{Second}, we show that networks instead learn a magnitude-aware frequency weighting that amplifies high-energy bins (the opposite of PHAT normalization), confirmed causally via single-bin masking. \textit{Third}, we demonstrate that PHAT whitening acts as an information bottleneck: removing it from both classical and neural GCC pipelines improves performance under additive noise, while end-to-end networks that retain full magnitude information outperform all fixed-weighting methods across conditions.
\section{Method}
\label{sec:method}

\subsection{Network architectures}
\label{sec:arch}

We select three architectures with systematically varying frequency connectivity. All networks receive the same narrowband observation vector. For each frequency bin $k$, we construct a 4-dimensional input:
\begin{equation}
\begin{split}
  \mathbf{t}[k] = \bigl[&\mathrm{Re}(X_1[k]),\, \mathrm{Im}(X_1[k]),\\
                         &\mathrm{Re}(X_2[k]),\, \mathrm{Im}(X_2[k])\bigr]
\end{split}
\label{eq:token}
\end{equation}
where $X_1[k]$ and $X_2[k]$ are the complex STFT coefficients of the two microphone channels at frequency bin $k$, and $\mathrm{Re}(\cdot)$ and $\mathrm{Im}(\cdot)$ denote real and imaginary parts. This representation is chosen so that the cross-power spectrum $G_{12}[k] = X_1[k] \cdot X_2^*[k]$, with $(\cdot)^*$ denoting complex conjugation, is a bilinear function of $\mathbf{t}[k]$, computable by a single feedforward layer, matching the first algorithmic step of GCC-PHAT by construction.

We train three architecturally distinct networks, where $d$ denotes hidden dimension and each ``layer'' refers to one feedforward block (MLP), convolutional block (CNN), or transformer block (Transformer). The \textbf{MLP-per-bin} (3 layers, $d{=}64$, 8.8k params) applies an independent shared-weight MLP to each token, isolating per-frequency computation: it can compute any function of $\mathbf{t}[k]$ but lacks the cross-frequency pathway needed for delay aggregation. We additionally train an MLP with $d{=}256$ (133k params) to control for capacity. The \textbf{1D-CNN} (4 layers, $d{=}64$, kernel 5, 129k params) applies convolutions along the frequency axis, providing local cross-frequency mixing that expands with depth. The \textbf{Transformer}~\cite{vaswani2017attention} (4 layers, $d{=}64$, 209k params) uses global self-attention with a learnable CLS token, enabling all-to-all frequency interaction from the first layer. All networks output a scalar delay estimate $\hat{\tau}$, where $\tau$ denotes the ground-truth inter-channel delay in samples and $\tau_{\max}{=}30$ samples bounds its range. The regression target is $\tau / \tau_{\max} \in [-1, 1]$, and we report mean absolute error $\mathrm{MAE} = |\hat{\tau} - \tau|$ in samples.

\subsection{Probing framework}
\label{sec:probing}

Our probing strategy (Fig.~\ref{fig:framework}) exploits the fact that GCC-PHAT's intermediate computations are fully specified, so each algorithmic step yields a ground-truth target. GCC-PHAT estimates TDOA in three stages.

\noindent First, the cross-power spectrum:
\begin{equation}
  G_{12}[k] = X_1[k] \cdot X_2^*[k].
  \label{eq:cross}
\end{equation}
Second, PHAT whitening normalizes $G_{12}$ to unit magnitude:
\begin{equation}
  G_{12}^{\mathrm{PHAT}}[k] = \frac{G_{12}[k]}{|G_{12}[k]|} = e^{j\angle G_{12}[k]}.
  \label{eq:phat}
\end{equation}
where $\angle$ denotes the complex phase.

\noindent Third, an inverse FFT of $G_{12}^{\mathrm{PHAT}}$ followed by peak-picking yields $\hat{\tau}$. If a network follows GCC-PHAT internally, then at some layer $l$ its representation $f_l(\mathbf{x})$ should encode $G_{12}$; if it further applies whitening, it should encode the phase $\angle G_{12}$ while discarding the magnitude $|G_{12}|$.

Following the probing classifiers framework~\cite{belinkov2022probing}, we test these hypotheses by training a probe $g: f_l(\mathbf{x}) \mapsto z$ from each layer's hidden states to each algorithmic target. We use Ridge regression ($\alpha{=}1.0$) as our linear probe. To quantify decodability, we measure the coefficient of determination:
\begin{equation}
  R^2 = 1 - \frac{\sum_i (y_i - \hat{y}_i)^2}{\sum_i (y_i - \bar{y})^2},
  \label{eq:rsquare}
\end{equation}
where $y_i$ is the ground-truth target value $z$ for test sample $i$, $\hat{y}_i$ is the probe prediction, and $\bar{y}$ is the mean of the targets. A high $R^2$ indicates that the target is \emph{linearly accessible} in the representation, i.e., available without additional nonlinear computation~\cite{alain2017understanding}. To verify that low linear $R^2$ reflects genuine absence rather than nonlinear encoding, we also train nonlinear MLP probes (2 hidden layers, dimension 128) as controls, following~\cite{hewitt2019designing}.

For each frequency bin $k$, we define four per-frequency targets:
\begin{align}
\text{cross}_\text{re}[k] &= \operatorname{Re}(G_{12}[k]), \quad
\text{cross}_\text{im}[k] = \operatorname{Im}(G_{12}[k]), \label{eq:probe_cross} \\
\text{phat}_\text{cos}[k] &= \cos(\frac{2\pi k\tau}{N_\text{FFT}}), \quad
\text{phat}_\text{sin}[k] = \sin(\frac{2\pi k\tau}{N_\text{FFT}}), \label{eq:probe_phat}
\end{align}
where $N_\text{FFT}{=}256$ is the FFT length, and $\phatcos$ and $\phatsin$ are the theoretical pure-phase values that ideal PHAT whitening would produce in the noiseless case. Together, these targets form a diagnostic pair: if a network follows GCC-PHAT internally, both should be decodable; if it computes cross-power but stops short of whitening, only~\eqref{eq:probe_cross} should yield high $R^2$. We additionally probe the scalar $\tau$ from the global (CLS or mean-pooled) token at each layer.

Because $\phatcos$ measures the \emph{theoretical} noiseless PHAT, low decodability might simply reflect noise corruption rather than absence of whitening. To address this, we define two supplementary targets: the observed (noisy) PHAT phase $\operatorname{Re}(G_{12}[k]/|G_{12}[k]|)$ and the cross-power magnitude $|G_{12}[k]|$. If the network encodes both cross-power and magnitude separately, observed PHAT is linearly reconstructible as their ratio without the network ever performing explicit division.

\subsection{Attribution and causal masking}
\label{sec:attribution}

Probing reveals \emph{what} the network represents, but not \emph{how} it weights frequencies during aggregation; we therefore compute gradient-based frequency attribution~\cite{simonyan2014deep}:
\begin{equation}
  w[k] = \bigl\lVert \partial \hat{\tau}\, /\, \partial \mathbf{t}[k] \bigr\rVert_2,
  \label{eq:grad}
\end{equation}
yielding an effective importance weight per frequency bin, averaged over the test set. If the network has learned PHAT-like weighting, $w[k]$ should correlate with $1/|G_{12}[k]|$; if it favors high-energy bins, with $|G_{12}[k]|$. We quantify this via Pearson correlation $r$.

Gradient attribution is correlational~\cite{shen2025reliability}. To establish whether the highlighted bins \emph{causally} affect performance, we apply single-bin frequency masking~\cite{zeiler2014visualizing}: for each bin $k$, we zero out the input token $\mathbf{t}[k]$ at test time and measure the change in MAE:
\begin{equation}
  \Delta\text{MAE}[k] = \text{MAE}_{\setminus k} - \text{MAE}_{\text{full}}.
  \label{eq:masking}
\end{equation}
If gradient attribution faithfully reflects the network's strategy, the masking sensitivity $\Delta\text{MAE}[k]$ should correlate with $w[k]$.

\begin{figure*}[t]
  \centering
  \includegraphics[width=0.84\textwidth]{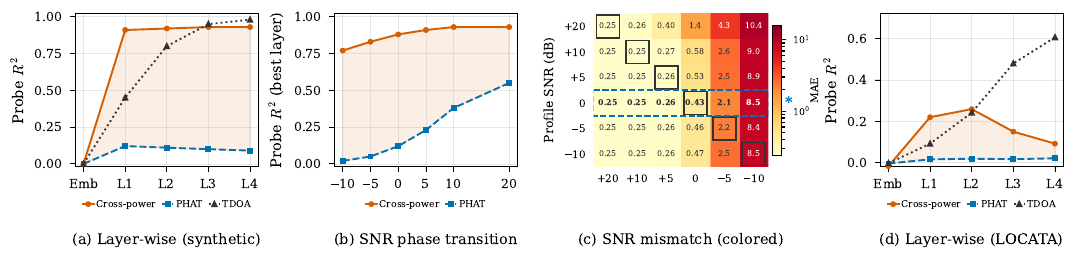}
  \caption{\textbf{Summary of core findings.} (a)~Layer-wise probing $R^2$ (Transformer, colored noise, 0\,dB): cross-power emerges at layer~1 and persists, while PHAT phase remains low throughout. (b)~SNR phase transition: cross-power $R^2$ is stable across SNR while PHAT collapses. (c)~SNR mismatch (colored noise): a single profile trained at 0\,dB (dashed box) generalizes across all test SNRs; squares mark the matched-SNR diagonal. (d)~LOCATA~\cite{lollmann2018locata} real recordings: the same qualitative pattern holds despite lower absolute $R^2$.}
  \label{fig:core}
\end{figure*}

\newpage
\section{Experiments and Results}
\label{sec:experiments}

\subsection{Setup}
\label{ssec:setup}

All models are trained with AdamW, cosine schedule, Huber loss, and batch size 1024 for 120 epochs; the default condition is colored noise at $0$\,dB. Probes are trained on 3000 samples (70\%/30\% train/test split).

\subsubsection{Synthetic data}
Synthetic data provides exact ground-truth delays and precise control over SNR, noise type, and source signal. We generate dual-channel signals as
\begin{equation}
  x_1[n] = s[n] + w_1[n], \qquad x_2[n] = s[n - \tau] + w_2[n],
  \label{eq:signal}
\end{equation}
where $s[n]$ is a spectrally shaped noise source, $\tau \sim \mathrm{Uniform}[-\tau_{\max},
\tau_{\max}]$, and $w_i[n]$ are independent white or colored ($1/f$)
noise processes at a specified SNR.
Signals are transformed via a
256-point STFT (hop size 128), yielding $F = 129$ frequency bins
per channel. For each condition we generate 50k training and 5k
validation samples with fixed seeds.

\subsubsection{Validation data}
To verify generalization beyond synthetic conditions, we progressively increase realism along two axes.

\textbf{Speech with simulated channels:} we replace the shaped-noise source with utterances from LibriSpeech~\cite{panayotov2015librispeech} and test under both direct-path delay and simulated reverberation using pyroomacoustics~\cite{scheibler2018pyroomacoustics} (ShoeBox rooms, $T_{60} \in \{0.2, 0.4, 0.6\}$\,s); delays and room geometry remain synthetically controlled.

\textbf{Real multi-channel recording:} we use the LOCATA
Challenge~\cite{lollmann2018locata} Task~1 (static speaker, DICIT
15-mic planar array), where delays arise from physical microphone positions and ground truth is computed geometrically from motion capture with inherent measurement error. We enumerate all $\binom{15}{2}{=}105$ mic
pairs across 3 recordings (224 pairs with
$0.5 < |\tau| \leq 30$), segmented at 16\,kHz into $T{=}256$
non-overlapping frames (72k total).

\begin{table}[t]
  \caption{Probing $R^2$ at best layer across synthetic (top) and validation (bottom) conditions.}
  \label{tab:crossarch}
  \centering
  \setlength{\tabcolsep}{3.5pt}
  \renewcommand{\arraystretch}{0.95}
  \begin{tabular}{l cccc}
    \toprule
    \textbf{Condition} & $\crossre$ & $\phatcos$ & $\boldsymbol{\tau}$ & \textbf{MAE} \\
    \midrule
    \multicolumn{5}{l}{\textit{Synthetic data}} \\
    \multicolumn{5}{l}{\quad\textit{(a) Cross-architecture (direct path, 0\,dB)}} \\
    \quad MLP-per-bin (8.8k)  & 0.84 & 0.11 & 0.89 & 2.32 \\
    \quad MLP-per-bin (133k)  & 0.90 & 0.12 & 0.96 & 1.58 \\
    \quad CNN (129k)       & 0.42 & 0.21 & 0.99 & 1.05 \\
    \quad Transformer (209k) & 0.94 & 0.12 & 0.98 & 1.53 \\
    \multicolumn{5}{l}{\quad\textit{(b) Reverberation (Transformer, 10\,dB)}} \\
    \quad $T_{60}{=}$0.2\,s & 0.80 & 0.19 & 0.86 & 3.23 \\
    \quad $T_{60}{=}$0.4\,s & 0.81 & 0.07 & 0.60 & 6.68 \\
    \quad $T_{60}{=}$0.6\,s & 0.62 & 0.02 & 0.44 & 8.42 \\
    \quad Mixed              & 0.75 & 0.07 & 0.59 & 6.52 \\
    \midrule
    \multicolumn{5}{l}{\textit{Validation data (Transformer)}} \\
    \multicolumn{5}{l}{\quad\textit{(c) Speech + simulated channel~\cite{panayotov2015librispeech}}} \\
    \quad Direct path (0\,dB) & 0.64 & 0.17 & 0.97 & 2.88 \\
    \quad Reverb (10\,dB) & 0.44 & 0.00 & 0.36 & 9.59 \\
    \multicolumn{5}{l}{\quad\textit{(d) Real multi-channel recording~\cite{lollmann2018locata}}} \\
    \quad LOCATA & 0.26 & 0.02 & 0.61 & 5.75 \\
    \bottomrule
  \end{tabular}
\end{table}

\subsection{Results and analysis}
\label{ssec:results}

We first identify \emph{which} GCC-PHAT steps networks replicate, then characterize \emph{what} alternative strategy they adopt, and finally \emph{what this implies} for pipeline design.

\noindent\textbf{Cross-power is replicated; PHAT whitening is bypassed.}\enspace
\label{sec:what}
We probe per-frequency hidden states for cross-power $\crossre$ and PHAT phase $\phatcos$ at each layer. Table~\ref{tab:crossarch} reports $R^2$ at the best layer.

Cross-power is linearly decodable after the first nonlinear layer in all three architectures (Table~\ref{tab:crossarch}a; CNN's lower $R^2$ reflects convolutional mixing; nonlinear probes recover $R^2{=}0.86$, Table~\ref{tab:controls}). The layer-wise trajectory (Fig.~\ref{fig:core}a) shows cross-power emerging at layer~1, with $\tau$ building progressively through deeper layers. This pattern holds across all conditions, with $\phatcos$ consistently near zero: reverberation (b), speech sources (c), real recordings (d; Fig.~\ref{fig:core}d), and SNR from $+20$ to $-10$\,dB (Fig.~\ref{fig:core}b). Under reverberation, $\tau$ decodability degrades with increasing $T_{60}$ ($R^2$: $0.86 \to 0.44$), but cross-power remains robust.

In contrast, $\phatcos$ remains poorly decodable in every condition of Table~\ref{tab:crossarch} ($R^2 \leq 0.21$). Table~\ref{tab:controls} consolidates the control analyses: nonlinear probes raise $\phatcos$ only to $R^2{=}0.23$, and increased model capacity (8.8k$\to$133k parameters) leaves it unchanged, confirming that PHAT whitening fails to emerge rather than being nonlinearly hidden. The network retains magnitude information because it encodes per-frequency signal reliability. Bins with higher $|G_{12}[k]|$ carry stronger signal energy relative to noise, making them more informative for delay estimation. PHAT discards exactly this information by normalizing all bins to unit magnitude. Probing for the observed PHAT phase $\operatorname{Re}(G_{12}/|G_{12}|)$ yields $R^2{=}0.69$, but this is a compositional artifact: the network separately encodes cross-power ($R^2{=}0.94$) and magnitude ($R^2{=}0.82$; Table~\ref{tab:controls}), from which observed PHAT is linearly reconstructible.

\begin{table}[t]
  \caption{Control probes (Transformer, colored noise, 0\,dB, best layer, unless noted). Nonlinear probes are 2-layer MLPs.}
  \label{tab:controls}
  \centering
  \setlength{\tabcolsep}{3pt}
  \begin{tabular}{l cc}
    \toprule
    \textbf{Probe target} & \textbf{Linear $R^2$} & \textbf{Nonlin.\ $R^2$} \\
    \midrule
    $\crossre$                                          & 0.94 & 0.94 \\
    $\phatcos$ (theoretical PHAT)                       & 0.12 & 0.23 \\
    Observed PHAT $\operatorname{Re}(G_{12}/|G_{12}|)$  & 0.69 & --   \\
    Magnitude $|G_{12}|$                                & 0.82 & --   \\
    $\crossre$ (CNN)                                    & 0.42 & 0.86 \\
    $\phatcos$ (MLP-per-bin, 8.8k\,/\,133k)             & 0.11\,/\,0.12 & -- \\
    \bottomrule
  \end{tabular}
\end{table}

\noindent\textbf{Networks adopt two-phase computation with magnitude-aware weighting.}\enspace
\label{sec:how}
The MLP-per-bin architecture processes each frequency independently, revealing the computational structure. Its hidden states faithfully encode per-bin cross-power (Phase~1), so $\tau$ remains highly \emph{probe}-decodable ($R^2{=}0.89$--$0.96$, Table~\ref{tab:crossarch}): the linear probe reads the full hidden state and aggregates per-bin features externally, much as GCC-PHAT recovers $\tau$ via the IFFT. The network itself, however, aggregates only through a uniform mean-pool head and lacks adaptive, data-dependent cross-frequency weighting (Phase~2). This deficit appears in task error rather than probe $R^2$: MAE 2.32 at 8.8k parameters versus 1.05 for the CNN (129k), and still 1.58 at matched capacity (133k). CNN and Transformer realize Phase~2 internally: the CNN's receptive field grows with depth, while the Transformer uses global attention from the first layer. This mirrors GCC-PHAT: the network substitutes a learned weighting for PHAT in Phase~1 and a learned aggregation for IFFT+argmax in Phase~2.

\begin{figure}[t]
  \centering
  \includegraphics[width=0.68\linewidth]{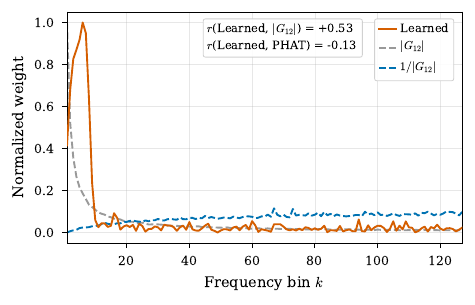}
  \caption{\textbf{Frequency-domain weighting profiles (Transformer, colored noise, 0\,dB).} The learned gradient-based weighting (orange) tracks cross-power magnitude $|G_{12}|$ (gray dashed) and is anti-correlated with PHAT weighting $1/|G_{12}|$ (blue dashed). All profiles normalized to $[0,1]$.}
  \label{fig:weighting}
\end{figure}

Gradient attribution~\eqref{eq:grad} reveals the nature of this weighting (Fig.~\ref{fig:weighting}): it is positively correlated with cross-power magnitude ($r{=}+0.53$) and anti-correlated with PHAT weighting ($r{=}{-}0.13$), resembling maximum-likelihood (ML)/Wiener weighting~\cite{brandstein2001microphone}. The pattern holds across SNR levels and for speech sources. Single-bin masking~\eqref{eq:masking} confirms this causally: $\Delta\text{MAE}[k]$ correlates with gradient attribution ($r{=}+0.94$) and magnitude ($r{=}+0.69$), and is anti-correlated with PHAT ($r{=}{-}0.66$).

\begin{figure}[t]
  \centering
  \includegraphics[width=\linewidth]{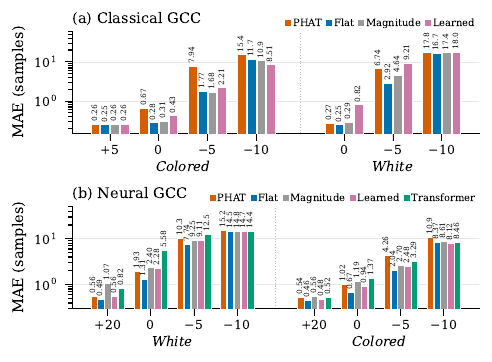}
  \caption{\textbf{GCC and NGCC benchmark (MAE in samples; exact values above bars).} (a)~Classical GCC with four weightings: Flat outperforms PHAT in all 7 conditions. (b)~Neural GCC (NGCC): Flat preprocessing outperforms PHAT in all 8 conditions shown (11 of 12 in the full SNR sweep, Sec.~\ref{ssec:results}). Lower is better; log scale.}
  \label{fig:gcc}
\end{figure}

\noindent\textbf{PHAT whitening is an information bottleneck for neural pipelines.}\enspace
\label{sec:implications}
We compare four GCC weighting functions $W[k]$ in $\hat{\tau} = \arg\max \mathrm{IFFT}(W[k]\, G_{12}[k])$: \textbf{PHAT} ($W{=}1/|G_{12}|$), \textbf{Flat} ($W{=}1$, plain cross-correlation), \textbf{Magnitude} ($W{=}|G_{12}|$), and \textbf{Learned} (the gradient-attribution profile of Eq.~\eqref{eq:grad}). In the classical setting (Fig.~\ref{fig:gcc}a), Flat outperforms GCC-PHAT in all seven conditions, PHAT is the worst of the four weightings in every colored-noise condition at ${\leq}\,0$\,dB, and the learned weighting achieves $44.8\%$ MAE reduction over PHAT at $-10$\,dB colored noise. A single learned profile generalizes across SNR levels (Fig.~\ref{fig:core}c): the spectral shape of magnitude weighting is SNR-invariant. This information loss is irrecoverable even with a downstream network. In the neural GCC (NGCC) setting~\cite{berg22_interspeech}, the weighted correlation is fed to a neural estimator instead of the argmax; all variants share the same 3-layer MLP and differ only in $W[k]$ (Fig.~\ref{fig:gcc}b). Flat preprocessing outperforms PHAT in 11 of 12 conditions across $\{{+}20,{+}10,{+}5,0,{-}5,{-}10\}$\,dB${}\times{}$\{white, colored\}, with up to $52\%$ MAE reduction; the sole exception is white noise at $+10$\,dB (Fig.~\ref{fig:gcc}b shows eight conditions). This directly challenges architectures built on GCC-PHAT preprocessing~\cite{berg22_interspeech}. On LOCATA real-world data, all classical methods approach chance level (MAE $13.9$--$17.4$), with GCC-PHAT best among them, consistent with PHAT suppressing reverberation artifacts~\cite{champagne1996performance, dibiase2001robust}; the Transformer achieves $2.4{\times}$ lower MAE ($5.75$) by retaining magnitude information and learning a data-adaptive weighting.

\section{Conclusions}
\label{sec:discussion}

By probing neural networks against GCC-PHAT's steps, we find that cross-power computation is a robust, architecture-shared inductive bias, while PHAT whitening consistently fails to emerge; networks instead learn magnitude-aware weighting that preserves the per-frequency reliability PHAT discards. In practice, removing PHAT from classical and neural pipelines improves performance under additive noise. However, our targets assume a single source, and whether these representations emerge in waveform-domain models is open. The $\tau$-decodability drop with $T_{60}$ (Table~\ref{tab:crossarch}b) further hints that reverberation may push networks toward features beyond GCC, e.g., implicit dereverberation, a target for future probing.

\section{Acknowledgments}
This study was carried out using the TSUBAME4.0 supercomputer at Institute of Science Tokyo.

\section{Generative AI Use Disclosure}
Claude Code (Anthropic) was used to assist with editing the experiment code. All experimental designs, analyses, and the manuscript text were produced and verified by the authors.

\bibliographystyle{IEEEtran}
\bibliography{refs}

@article{wang2025observability,
  author  = {Wang, Jiang and Kang, Yaozhong and Fu, Linya and Nakadai, Kazuhiro and Kong, He},
  title   = {Observability-Aware Active Calibration of Multisensor Extrinsics for Ground Robots via Online Trajectory Optimization},
  journal = {IEEE Sensors Journal},
  volume  = {25},
  number  = {17},
  pages   = {33022--33036},
  year    = {2025},
}

@article{knapp1976generalized,
  author  = {Knapp, Charles and Carter, G. Clifford},
  title   = {The Generalized Correlation Method for Estimation of Time Delay},
  journal = {IEEE Trans. Acoust., Speech, Signal Process.},
  volume  = {24},
  number  = {4},
  pages   = {320--327},
  year    = {1976},
}

@inproceedings{berg22_interspeech,
  author    = {Berg, Axel and O'Connor, Mark and {\AA}str{\"o}m, Kalle and Oskarsson, Magnus},
  title     = {Extending {GCC-PHAT} using Shift Equivariant Neural Networks},
  booktitle = {Proc. Interspeech 2022},
  pages     = {1791--1795},
  year      = {2022},
  doi       = {10.21437/Interspeech.2022-524},
}

@inproceedings{salvati21_interspeech,
  author    = {Salvati, Daniele and Drioli, Carlo and Foresti, Gian Luca},
  title     = {Time Delay Estimation for Speaker Localization Using {CNN}-Based Parametrized {GCC-PHAT} Features},
  booktitle = {Proc. Interspeech 2021},
  pages     = {1479--1483},
  year      = {2021},
  doi       = {10.21437/Interspeech.2021-988},
}

@inproceedings{alain2017understanding,
  author    = {Alain, Guillaume and Bengio, Yoshua},
  title     = {Understanding Intermediate Layers Using Linear Classifier Probes},
  booktitle = {Proc. ICLR Workshop},
  year      = {2017},
}

@article{belinkov2022probing,
  author  = {Belinkov, Yonatan},
  title   = {Probing Classifiers: Promises, Shortcomings, and Advances},
  journal = {Computational Linguistics},
  volume  = {48},
  number  = {1},
  pages   = {207--219},
  year    = {2022},
}

@inproceedings{nanda2023progress,
  author    = {Nanda, Neel and Chan, Lawrence and Lieberum, Tom and Smith, Jess and Steinhardt, Jacob},
  title     = {Progress Measures for Grokking via Mechanistic Interpretability},
  booktitle = {Proc. ICLR},
  year      = {2023},
}

@article{monga2021algorithm,
  author  = {Monga, Vishal and Li, Yuelong and Eldar, Yonina C.},
  title   = {Algorithm Unrolling: Interpretable, Efficient Deep Learning for Signal and Image Processing},
  journal = {IEEE Signal Process. Mag.},
  volume  = {38},
  number  = {2},
  pages   = {18--44},
  year    = {2021},
}

@inproceedings{lollmann2018locata,
  author    = {L{\"o}llmann, Heinrich W. and Evers, Christine and Schmidt, Alexander and Mellmann, Heinrich and Barfuss, Hendrik and Naylor, Patrick A. and Kellermann, Walter},
  title     = {The {LOCATA} Challenge Data Corpus for Acoustic Source Localization and Tracking},
  booktitle = {Proc. IEEE Sensor Array and Multichannel Signal Processing Workshop (SAM)},
  pages     = {410--414},
  year      = {2018},
}

@inproceedings{scheibler2018pyroomacoustics,
  author    = {Scheibler, Robin and Bezzam, Eric and Dokmanić, Ivan},
  title     = {Pyroomacoustics: A {P}ython Package for Audio Room Simulation and Array Processing Algorithms},
  booktitle = {Proc. IEEE ICASSP},
  pages     = {351--355},
  year      = {2018},
}

@inproceedings{pertila2019tdoa,
  author    = {Pertil{\"a}, Pasi and Parviainen, Mikko},
  title     = {Time Difference of Arrival Estimation of Speech Signals Using Deep Neural Networks with Integrated Time-frequency Masking},
  booktitle = {Proc. IEEE ICASSP},
  pages     = {436--440},
  year      = {2019},
}

@article{champagne1996performance,
  author  = {Champagne, Beno{\^\i}t and B{\'e}dard, St{\'e}phane and St{\'e}phenne, Alex},
  title   = {Performance of Time-Delay Estimation in the Presence of Room Reverberation},
  journal = {IEEE Trans. Speech Audio Process.},
  volume  = {4},
  number  = {2},
  pages   = {148--152},
  year    = {1996},
}

@article{ianniello1982time,
  author  = {Ianniello, Joseph P.},
  title   = {Time Delay Estimation via Cross-Correlation in the Presence of Large Estimation Errors},
  journal = {IEEE Trans. Acoust., Speech, Signal Process.},
  volume  = {30},
  number  = {6},
  pages   = {998--1003},
  year    = {1982},
}

@incollection{dibiase2001robust,
  author    = {DiBiase, Joseph H. and Silverman, Harvey F. and Brandstein, Michael S.},
  title     = {Robust Localization in Reverberant Rooms},
  booktitle = {Microphone Arrays: Signal Processing Techniques and Applications},
  editor    = {Brandstein, Michael S. and Ward, Darren B.},
  pages     = {157--180},
  publisher = {Springer},
  year      = {2001},
}

@article{grumiaux2022survey,
  author  = {Grumiaux, Pierre-Amaury and Kitić, Srđan and Girin, Laurent and Guérin, Alexandre},
  title   = {A Survey of Sound Source Localization with Deep Learning Methods},
  journal = {J. Acoust. Soc. Am.},
  volume  = {152},
  number  = {1},
  pages   = {107--151},
  year    = {2022},
}

@inproceedings{vaswani2017attention,
  author    = {Vaswani, Ashish and Shazeer, Noam and Parmar, Niki and Uszkoreit, Jakob and Jones, Llion and Gomez, Aidan N. and Kaiser, {\L}ukasz and Polosukhin, Illia},
  title     = {Attention Is All You Need},
  booktitle = {Proc. NeurIPS},
  pages     = {5998--6008},
  year      = {2017},
}

@inproceedings{zhang2008phat,
  author    = {Zhang, Cha and Flor{\^e}ncio, Dinei and Zhang, Zhengyou},
  title     = {Why Does {PHAT} Work Well in Low Noise, Reverberative Environments?},
  booktitle = {Proc. IEEE ICASSP},
  pages     = {2565--2568},
  year      = {2008},
}

@inproceedings{voita2020information,
  author    = {Voita, Elena and Titov, Ivan},
  title     = {Information-Theoretic Probing with Minimum Description Length},
  booktitle = {Proc. EMNLP},
  pages     = {183--196},
  year      = {2020},
}

@inproceedings{cho2023evidence,
  author    = {Cho, Cheol Jun and Wu, Peter and Mohamed, Abdelrahman and Anumanchipalli, Gopala K.},
  title     = {Evidence of Vocal Tract Articulation in Self-Supervised Learning of Speech},
  booktitle = {Proc. IEEE ICASSP},
  pages     = {1--5},
  year      = {2023},
}

@inproceedings{conneau2018probing,
  author    = {Conneau, Alexis and Kruszewski, German and Lample, Guillaume and Barrault, Lo{\"i}c and Baroni, Marco},
  title     = {What You Can Cram into a Single \$\&!\#* Vector: Probing Sentence Embeddings for Linguistic Properties},
  booktitle = {Proc. ACL},
  pages     = {2126--2136},
  year      = {2018},
}

@inproceedings{hewitt2019designing,
  author    = {Hewitt, John and Liang, Percy},
  title     = {Designing and Interpreting Probes with Control Tasks},
  booktitle = {Proc. EMNLP},
  pages     = {2733--2743},
  year      = {2019},
}

@inproceedings{shen2025reliability,
  author    = {Shen, Gaofei and Mohebbi, Hosein and Bisazza, Arianna and Alishahi, Afra and Chrupa{\l}a, Grzegorz},
  title     = {On the Reliability of Feature Attribution Methods for Speech Classification},
  booktitle = {Proc. Interspeech},
  year      = {2025},
}

@inproceedings{choi2024phonetic,
  author    = {Choi, Kwanghee and Pasad, Ankita and Nakamura, Tomohiko and Fukayama, Satoru and Livescu, Karen and Watanabe, Shinji},
  title     = {Self-Supervised Speech Representations are More Phonetic than Semantic},
  booktitle = {Proc. Interspeech},
  pages     = {4578--4582},
  year      = {2024},
  doi       = {10.21437/Interspeech.2024-1157},
}

@inproceedings{simonyan2014deep,
  author    = {Simonyan, Karen and Vedaldi, Andrea and Zisserman, Andrew},
  title     = {Deep Inside Convolutional Networks: Visualising Image Classification Models and Saliency Maps},
  booktitle = {Proc. ICLR Workshop},
  year      = {2014},
}

@inproceedings{zeiler2014visualizing,
  author    = {Zeiler, Matthew D. and Fergus, Rob},
  title     = {Visualizing and Understanding Convolutional Networks},
  booktitle = {Proc. European Conference on Computer Vision (ECCV)},
  pages     = {818--833},
  year      = {2014},
}

@book{brandstein2001microphone,
  editor    = {Brandstein, Michael and Ward, Darren},
  title     = {Microphone Arrays: Signal Processing Techniques and Applications},
  publisher = {Springer},
  year      = {2001},
}

@article{politis2021overview,
  author  = {Politis, Archontis and Mesaros, Annamaria and Adavanne, Sharath and Heittola, Toni and Virtanen, Tuomas},
  title   = {Overview and Evaluation of Sound Event Localization and Detection in {DCASE} 2019},
  journal = {IEEE/ACM Trans. Audio, Speech, Lang. Process.},
  volume  = {29},
  pages   = {684--698},
  year    = {2021},
}

@inproceedings{panayotov2015librispeech,
  author    = {Panayotov, Vassil and Chen, Guoguo and Povey, Daniel and Khudanpur, Sanjeev},
  title     = {{LibriSpeech}: An {ASR} Corpus Based on Public Domain Audio Books},
  booktitle = {Proc. IEEE ICASSP},
  pages     = {5206--5210},
  year      = {2015},
}

@article{hu2025pseldnets,
  author  = {Hu, Jinbo and Cao, Yin and Wu, Ming and Kang, Fang and Yang, Feiran and Wang, Wenwu and Plumbley, Mark D. and Yang, Jun},
  title   = {{PSELDNets}: Pre-Trained Neural Networks on a Large-Scale Synthetic Dataset for Sound Event Localization and Detection},
  journal = {IEEE/ACM Trans.\ Audio, Speech, Lang.\ Process.},
  volume  = {33},
  pages   = {2845--2860},
  year    = {2025},
}

\end{document}